\pdfoutput=1
\documentclass[fleqn,10pt]{wlscirep}
\usepackage[utf8]{inputenc}
\usepackage[T1]{fontenc}

\newcommand{\tb}{{t_{(\rm TB)}}}
\newcommand{\ltb}{{\ell_{(\rm TB)}}}
\newcommand{\chrh}{{\tilde{\eta}}}
\newcommand{\opth}{\eta^{({\rm opt})}}
\newcommand{\optphi}{\phi^{({\rm opt})}}
\newcommand{\af}{a^{({\rm F})}}
\newcommand{\ab}{a^{({\rm B})}}

\title{Optimizing hospital distribution across districts for reducing tuberculosis fatalities}

\author[1]{Mi Jin Lee}
\author[2]{Kanghun Kim}
\author[3]{Junik Son}
\author[1, *]{Deok-Sun Lee}

\affil[1]{Department of Physics, Inha University, Incheon 22212, Korea}
\affil[2]{Financial Engineering Team, Mertiz Securities, Seoul 07326, Korea}
\affil[3]{Chungnam National University Hospital, Daejeon 35015, Korea}

\affil[*]{deoksun.lee@inha.ac.kr}

\begin{abstract}
The spatial distributions of diverse facilities is often understood in terms of the optimization of the commute distance or the economic profit. Incorporating more  general objective functions into such optimization framework may be useful, helping the policy decisions to meet various social and economic demands. As an example, we consider how  hospitals should be distributed to minimize the total fatalities of tuberculosis (TB). The empirical data of Korea shows that the fatality rate of TB in a district  decreases with the areal density of hospitals, implying their correlation and  the possibility of reducing the nationwide fatalities by adjusting the hospital distribution across districts.  Approximating the fatality rate by the probability of a patient not to visit a hospital in her/his residential district for the duration period of TB and evaluating the latter probability in the random-walk framework, we obtain the fatality rate as an exponential function of the hospital density with a characteristic constant related to each district's effective lattice constant estimable empirically. This leads us to  the optimal hospital distribution which  finds the hospital density in a district to be  a logarithmic function of the rescaled patient density. The total fatalities is reduced by 13\% with this optimum. The current hospital density deviates from the optimized one in different manners from district to district, which is analyzed in the proposed model framework. The assumptions and limitations of our study are also discussed.
\end{abstract}
\begin{document}

\flushbottom
\maketitle

\thispagestyle{empty}

%\noindent Please note: Abbreviations should be introduced at the first mention in the main text ??no abbreviations lists. Suggested structure of main text (not enforced) is provided below.

%%%%%%%%%%%%%%%%%%%%%%%%%%%%%%%%%%%%%%%%%%%%%%%%%%%%%%%
%%% Main 
%%%%%%%%%%%%%%%%%%%%%%%%%%%%%%%%%%%%%%%%%%%%%%%%%%%%%%%

%%%%%%%%%%%%%%%%%%%%%%%%%%%%%%%%%%%%%%%%%%%%%%%%%%%%%%%
\section*{Introduction}

Complex systems are organized, by evolution or design,  to satisfy the optimization conditions including the minimization of the traveling time in the transportation system~\cite{youn2008} and the maximization of the  stability of the airline networks~\cite{wuellner2010},  the resilience of the power-grid system~\cite{witthaut2012, lee2017}, and the growth rate of cellular networks~\cite{motter2008}. Likewise, the locations of facilities are expected to be subject to various optimization conditions~\cite{OWEN1998423, 10003623242, CALVO1973407}.  Despite the complexity of the facility location decisions~\cite{doi:10.1137/0213014,CURRENT1990295},  the empirically observed distributions of facilities often show simple and universal features, revealing the nature of the underlying optimization problem. Most remarkably,  the spatial density of facilities scaling with the population density~\cite{science,bunge,newman} with  exponent $2/3$ or $1$ implies that they are distributed to minimize the social opportunity cost such as the commute distance  or to maximize the economic profit depending on the distribution of available customers~\cite{pnas}.  

For coping with diverse social or economic demands in real-world applications, the objective function in the facility distribution optimization may need to be expanded beyond the commute distance or profit. Towards developing such a  general theory, here we consider as an example the problem of distributing hospitals across districts to minimize the total fatalities of tuberculosis (TB) by using the empirical data of Korea. While the chemotherapy for TB  is well established, showing a success rate as high as 85\% on average~\cite{who}, TB spreads annually to about 10 million patients, being  a major cause of death worldwide~\cite{who, 2015report}. In Korea, the incidence of TB is 77 per 100000 as of 2016, which is high compared with other developed countries, e.g., the member countries of the Organization for Economic Cooperation and Development~\cite{doi:10.4178/epih.e2018036}. Patients with TB can be cured if they are diagnosed and get treatment timely. Visiting a hospital and taking drugs for about 6 months are necessary for the full recovery from TB~\cite{cdc}, which may not be easy from the patients' perspective.  Therefore, the accessibility of local hospitals and the well-trained attending staff providing consistent treatment and care should be crucial for the treatment of TB~\cite{2015report, HOPEWELL2006710}, which is recognized also in the reports of the World Health Organization~\cite{who}.  The correlation between the hospital distribution and the fatality rate of TB in a district is indeed identified in the Korea TB data-sets which we will analyze in the present study; The fatality rate in a district tends to decrease as the areal density of hospitals therein increases, which is an important point demanding a quantitative explanation and leads us to expect that relocating hospitals across districts may reduce the total  fatalities of TB nationwide.
 
The optimal distribution of hospitals across districts minimizing the total TB fatalities depends on the concrete form of the fatality rate as a function of the hospital density, which is, however, unknown; The empirically observed negative dependence cannot give this information, as districts are different not only in the hospital density but also in various other properties such as area or population. To address the district-dependent fatality rate, we take a modeling approach, in which  the fatality rate is assumed to be identical to the probability of a patient {\it not} to visit a hospital and get the medical treatment for the duration period of TB. This is motivated by the expectation that a patient is very likely to be cured once she/he gets a proper treatment in a hospital, given  the high success rate of the TB chemotherapy equally applicable to all districts. In this framework, the fatality rate turns out to be an exponentially decaying function of  the hospital density, and we are able to derive the optimal hospital densities in all districts the collection of which decreases the total fatalities of TB by 13\% from the current value. The predicted optimal hospital density is given by a logarithmic function of the rescaled patient density. Our results delineate an analytic approach to the facility optimization problem under a general objective function, leaving space for improvement and further generalization to be discussed.
%%%%%%%%%%%%%%%%%%%%%%%%%%%%%%%%%%%%%%%%%%%%%%%%%%%%%%%

\section*{Results}
\label{sec:results}

%%%%%%%%%%%%%%%%%%%%%%%%%%%%%%%%%%%%%%%%%%%%%%%%%%%%%%%
\subsection*{TB  fatality rate and hospital density: Empirical data}
\label{subsec:relation}

%%%%%%%%%% Figure 1: Empirical data for Fatality and hospital density 
\begin{figure}
\centering
\includegraphics[width=1.0\linewidth]{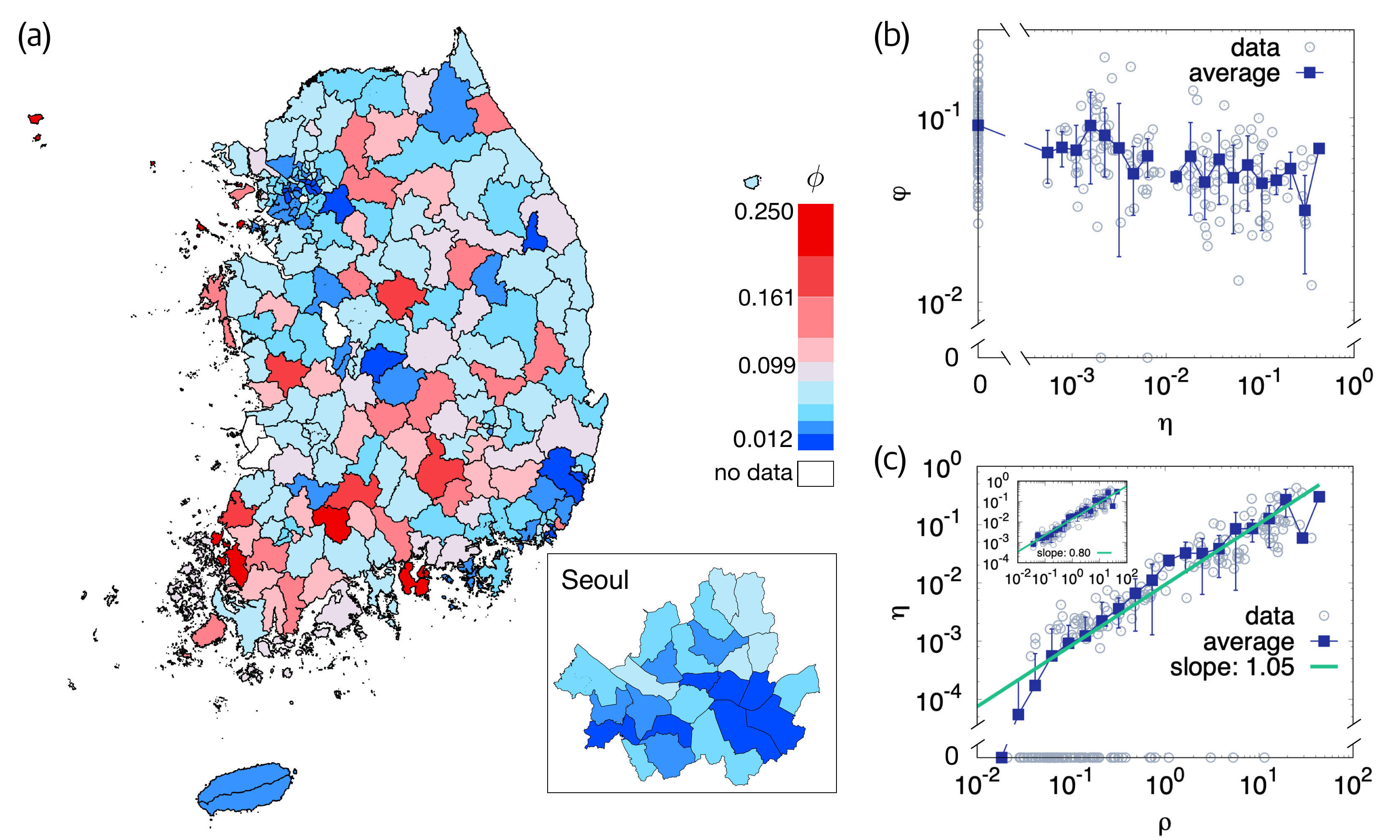}
\centering
\caption{
Distribution and relations of the TB  fatality rate $\phi$, the areal density of hospitals $\eta$, and the patient density $\rho$ in Korean districts at the level of Gu, Gun, and Si. The unit of $\eta$ and $\rho$ is km$^{-2}$.
({\bf a}) The TB  fatality rate $\phi$ is represented by color in 228 districts.  Seoul, the capital city, has 25 Gu's and is shown separately. 
({\bf b}) Fatality rate $\phi$ versus hospital density $\eta$.  Open circles are the raw data for all districts and filled squares represent the average fatality rate for each given  hospital density with the standard deviations as errorbars. The Pearson correlation coefficient  is $-0.26$ with P-value $0.000070$ for all districts and $-0.25$ with P$=0.0025$ for the districts with both $\eta$ and $\phi$ non-zero. 
({\bf c}) Hospital density $\eta$ versus patient density $\rho$. As in (b), open circles and filled squares represent the raw data and the average value, respectively. The solid line fits the averaged data and its slope is $1.05 \pm 0.07$. Inset: The same plot for the districts with $\eta>0$. The fitting line has slope $0.80 \pm 0.045$.
}
\label{fig:relations}
\end{figure}
%%%%%%%%%%%%%%%%%%%%%%%%%%%%%%%%%%%%%%%%%%%%%

 The incidence and mortality of TB  are well recorded in Korea. In Statistics Korea~\cite{kosis}, we obtain for district $i=1,2,\ldots, I=228$ in year 2014 the number of the newly reported TB  patients $N_i$, the number of dead TB  patients (fatalities) $D_i$, the number of private general hospitals $H_i$, and the area $A_i$. Here ``district" includes three distinct units for administrative division, Gu, Gun, and Si, with the population ranging from $10^4$ to $10^6$ and smaller than the metropolitan cities. 

We are interested in the fatality rate $\phi_i$ of TB, defined as the ratio  of the number of dead TB patients to the number of new TB  patients reported for one year in each district $i$, 
\begin{equation}
\phi_i \equiv {D_i \over N_i}.
\label{eq:phi_def}
\end{equation}
It is quite different from district to district, ranging between $0.01$ and $0.25$, as shown graphically in Fig.~\ref{fig:relations}(a). What drives such difference in the TB  fatality rate?  Taking regularly medical treatments and examinations in hospitals may be the most important for curing TB, which is available in the easy-to-frequently-access medical environment established in the local community.  Therefore  a difference in the abundance and accessibility of hospitals in the patients' residential districts will be a major factor giving rise to such variation of the fatality rate with district. In this light, we investigate the relation between the fatality rate $\phi_i$ and the areal hospital density 
\begin{equation}
\eta_i \equiv {H_i \over A_i}, 
\label{eq:eta_def}
\end{equation}
in unit of $\textrm{km}^{-2}$. In Fig.~\ref{fig:relations}(b) $\phi_i$ tends to decrease with $\eta_i$; The larger the hospital density is, the smaller the fatality rate is. This correlation is significant with P $<10^{-4}$. Yet  the dependence does not look so strong as expected. This will be shown to be due to  that the fatality rate of a district may depend not only on the hospital density but also on other characteristics.

What principle underlies the current spatial distribution of hospitals? The scaling behavior with respect to the patient density has hinted at the answer~\cite{pnas}. The hospital density scales with the areal TB  patient density $\rho_i\equiv {N_i \over A_i}$  as 
\begin{equation}
\eta_i \sim \rho_i^\alpha,
\label{eq:scal_eta}
\end{equation}
in which $\alpha= 1.05\pm 0.07$ when all districts are included, and $\eta=0.80\pm 0.045$ when the districts having no private general hospital are excluded [Fig.~\ref{fig:relations}(c)]. Many other properties also scale with respect to the patient density. The patient density is almost linearly related to  the population density $\rho'_i = P_i/A_i$ with $P_i$ the number of people living in district $i$ [Fig.~\ref{fig:scal_other}].  The exponent $\alpha$ for the hospitals in United States is  close to $1$, rather than  $2/3$~\cite{pnas}. These results suggest that the profit maximization affects the hospital distribution. For self-containment, let us sketch the corresponding optimization calculations. The sum of the economic profits of all hospitals distributed across $I'$ districts in a country is given by 
\begin{equation}
E_{\rm profit} = \sum_{i=1}^{I'} H_i \, \omega\left({N_i \over H_i}\right) = \sum_i A_i \, \eta_i \, \omega\left({\rho_i \over \eta_i}\right)
\label{eq:Eprofit}
\end{equation}
with $\omega(x)$  the expected profit of a single hospital having $x$ patients available. On the other hand, the sum of the social costs, such as the travel distances, of patients is given by 
\begin{equation}
E_{\rm cost} = \sum_{i=1}^{I'} N_i \, \psi\left({A_i \over H_i}\right) = \sum_i A_i \, \rho_i \, \psi\left({1 \over \eta_i}\right)
\label{eq:Ecost}
\end{equation}
with $\psi(x) = x^{1/2}$ being the expected travel distance of a patient residing in a district of $x$ area per hospital. Then, for a fixed total number of hospitals 
\begin{equation}
H_{\rm total}= \sum_{i=1}^{I'} H_i = \sum_I A_i \eta_i,
\label{eq:Htotal}
\end{equation}
one finds, by solving ${\partial  E_{\rm profit} \over \partial \eta_i} = 0$,  $E_{\rm profit}$ to be maximized when ${\rho_i \over \eta_i} = {\rm const.}$, corresponding to $\alpha=1$, and by solving ${\partial  E_{\rm cost} \over \partial \eta_i} = 0$,  $E_{\rm cost}$ to be minimized when
${\rho_i \over \eta_i^{3/2}} = {\rm const.}$, corresponding to $\alpha=2/3$~\cite{pnas}.

Our question is then whether the current hospital distribution, seemingly maximizing the economic profit, is the best also for minimizing the total fatalities of TB
\begin{equation}
E_{\rm fatalities}  = \sum_{i=1}^{I'} D_i = \sum_i N_i \, \phi_i.
\label{eq:Efatalities}
\end{equation}
Can $E_{\rm fatalities}$ be reduced by the redistribution of hospitals across district, i.e., some change of $\{\eta_i\}$?  To answer this, we should formulate  the total fatalities in Eq.~(\ref{eq:Efatalities}) as the objective function and minimize it with respect to the hospital density for the given total number of hospitals in Eq.~(\ref{eq:Htotal}). The fatality rate $\phi_i$  should be some function of the hospital density $\eta_i$. If the optimal hospital densities $\{\opth_i\}$ are obtained by this optimization computation, we will be able to evaluate the quality of the current spatial distribution of hospitals regarding its capacity of TB treatment. Also we will see immediately how to redistribute the hospitals to reduce the TB  fatalities.  In the present study we do not consider a variation in the numbers of TB  patients $\{N_i\}$ but take them for given; The onset and spreading of the TB  or a general epidemic disease depend strongly on the topology of human contact networks and the infection rate, which is another important research topic and has been studied extensively~\cite{PhysRevLett.86.3200,PhysRevE.99.032309}.
%%%%%%%%%%%%%%%%%%%%%%%%%%%%%%%%%%%%%%%%%%%%%%%%%%%%%%%

%%%%%%%%%%%%%%%%%%%%%%%%%%%%%%%%%%%%%%%%%%%%%%%%%%%%%%%
\subsection*{Fatality rate as a function of hospital density: Model}
\label{subsec:model}

The empirical fatality rate $\phi_i$ in Eq.~(\ref{eq:phi_def}) can be considered as the probability of a TB patient to die, losing the opportunity to get proper medical treatment in time. Our idea is to approximate the latter by the probability that a patient does not visit any hospital in her/his residential district for a given period $\tb=3$ years, the empirically reported period of TB  duration from onset to either cure or death~\cite{3yr}. In this model framework, it determines the fate of a TB patient whether she/he visits a hospital or not for the period of $\tb$. The patient will recover if yes, but will be dead otherwise. One can see that this is a trapping problem~\cite{hughesbook} from the viewpoint of a patient; Once a patient (walker) reaches a hospital (trap), she loses the status of a patient (absorbed at the trap).  The probability of a walker to {\it survive} during a given number of steps corresponding to $\tb$ in this trapping problem is translated into the {\it fatality} rate of a TB  patient in reality. 

Suppose that $H$ traps are uniformly and independently distributed in a two-dimensional Euclidean lattice of $L\times L$ sites and that a walker walks around the region, who disappears on reaching any one of the traps. Then the probability of the walker to survive (not to reach any of the traps) after $\tau$ steps is given by  
\begin{equation}
\phi = \langle (1-\lambda)^{S(\tau)}\rangle,
\label{eq:phi_exact}
\end{equation} 
where $\lambda = \frac{H}{L^2}$ is the density of traps and $S(\tau)$ is the number of distinct sites visited up to $\tau$ steps.   $\langle \cdots \rangle$ represents the average over different realizations of walks. In the limit $\left|\log (1-\lambda) {\sigma^2_{S(\tau)} \over \langle S(\tau)\rangle}\right| \ll 1$ reachable when  the trap density is sufficiently low or the number of steps is small enough,  the survival probability  $\phi$ can be approximated in terms of the first cumulant of the probability distribution of $S$ as~\cite{rw2} 
\begin{equation}
\phi = e^{-\lambda \langle S(\tau)\rangle},
\label{eq:phi_S}
\end{equation}
which is the exponential function of the trap density $\lambda$.  It seems that Eq.~(\ref{eq:phi_S}) allows us to relate the hospital density and the fatality rate. However the dimensionless quantities $\lambda$ and $\langle S(\tau)\rangle$ are not directly available. In random walks in two dimensions, the expected number of distinct visited sites is known to be~\cite{hughesbook}   
\begin{equation}
\langle S(\tau)\rangle\propto {\tau \over \log \tau},
\end{equation}
which is inserted into Eq.~(\ref{eq:phi_S}) to give 
\begin{equation}
\phi  =   \exp\left(-c \frac{\tau}{\log \tau} \lambda \right)
\label{eq:surv}
\end{equation}
with the coefficient $c=3.5/1.13^2$ known numerically~\cite{rw1}.  In the opposite limit $\left|\log (1-\lambda) {\sigma^2_{S(\tau)} \over \langle S(\tau)\rangle}\right| \gg 1$, the survival of the walker is governed by the probability of  a  large trap-free region to be formed, which leads to a stretched exponential form  $\log \phi \sim \sqrt{\lambda \tau}$~\cite{rw1,rw3,doi:10.1063/1.443832}. 

The exponential decay of $\phi$ with $\lambda$ in Eq.~(\ref{eq:phi_S}) holds when the hospital density is sufficiently low. The randomness of the mobility pattern is assumed in obtaining Eq.~(\ref{eq:surv}). We should remark that the human mobility pattern revealed by tracing the travel routes of bank notes~\cite{v0} or the mobile phone records~\cite{gonzalez2008} displays deviation from random walk; The radius of gyration of individual trajectories grows logarithmically with time~\cite{gonzalez2008}, in contrast to the square-root scaling in the conventional random walk, and such slow diffusion is known to arise under the memory effect~\cite{Song1018,Choi_2012,PhysRevE.93.052310} or the spatial quenched disorder~\cite{hughesbook}.  The assumption we make about the human mobility pattern is that the {\it coarse-grained} trajectories of individuals on the time scale of $\tb=3$ years, much longer than the previous studies, show the survival probability given in Eq.~(\ref{eq:surv}) like random walks. The coarse-grained trajectory is obtained by neglecting the spots swiftly passed by and connecting the remaining notable places which an individual visits and stays for a while in, such as her/his house, workplace, parks, stores, banks, oil stations, and hospitals. In our model, we are interested  in whether a hospital is included in the list of such notable places. We cannot check directly the validity of Eqs.~(\ref{eq:phi_S}) and (\ref{eq:surv}), however, we will present indirect evidence that they are reasonable assumptions. 
%%%%%%%%%%%%%%%%%%%%%%%%%%%%%%%%%%%%%%%%%%%%%%%%%%%%%%%

%%%%%%%%%%%%%%%%%%%%%%%%%%%%%%%%%%%%%%%%%%%%%%%%%%%%%%%
\subsection*{Lattice constant and dimensionless quantities}
\label{subsec:lattice}

To relate the survival probability in the 2D trapping problem to the fatality rate of TB, we need to convert the empirical data into dimensionless ones of Eq.~(\ref{eq:surv}). To this end, we discretize the region of each district $i$ by introducing the lattice constant $a_i$, corresponding to the typical length of one single step or the  average distance between adjacent notable places appearing in the coarse-grained trajectories. Then the district is represented by the $L_i\times L_i$  Euclidean lattice  with $L_i = \sqrt{A_i  \over a_i^2}$, for which  the areal hospital density $\eta_i=H_i/A_i$ is converted to the dimensionless hospital density $\lambda_i$ as
\begin{equation}
\lambda_i = {H_i \over L_i^2} = {H_i \over {A_i \over a_i^2}}=a_i^{2}\eta_i.
\label{eq:dimensionless_lambda}
\end{equation}
Let $\ltb$ be the typical travel distance of an individual  for $\tb=3$ years. Then the number of steps taken in her/his coarse-grained trajectory for $\tb$ in a district $i$ will be given by
\begin{equation}
\tau_i = \frac{\ltb}{a_i}.
\label{eq:dimensionless_tau}
\end{equation}
Plugging Eqs.~(\ref{eq:dimensionless_lambda}) and~(\ref{eq:dimensionless_tau}) into Eq.~(\ref{eq:surv}), we find the fatality rate represented as 
\begin{equation}
\phi_i = \exp \left( -\frac{\eta_i}{\chrh_i}\right)
\label{eq:fatality_rate}
\end{equation}
with the characteristic hospital density $\chrh_i$ given by
\begin{equation}
\chrh_i = \left( c\frac{\tau_i}{\log \tau_i} a_i^2 \right)^{-1} = \left( c\frac{\ltb a_i }{\log \left( {\ltb \over a_i} \right)} \right)^{-1}.
\label{eq:etatilde}
\end{equation}

error%%%%%%%%%%%%%%%%% Figure 2: Model prediction %%%%%%
\begin{figure}
\includegraphics[width=0.7\linewidth]{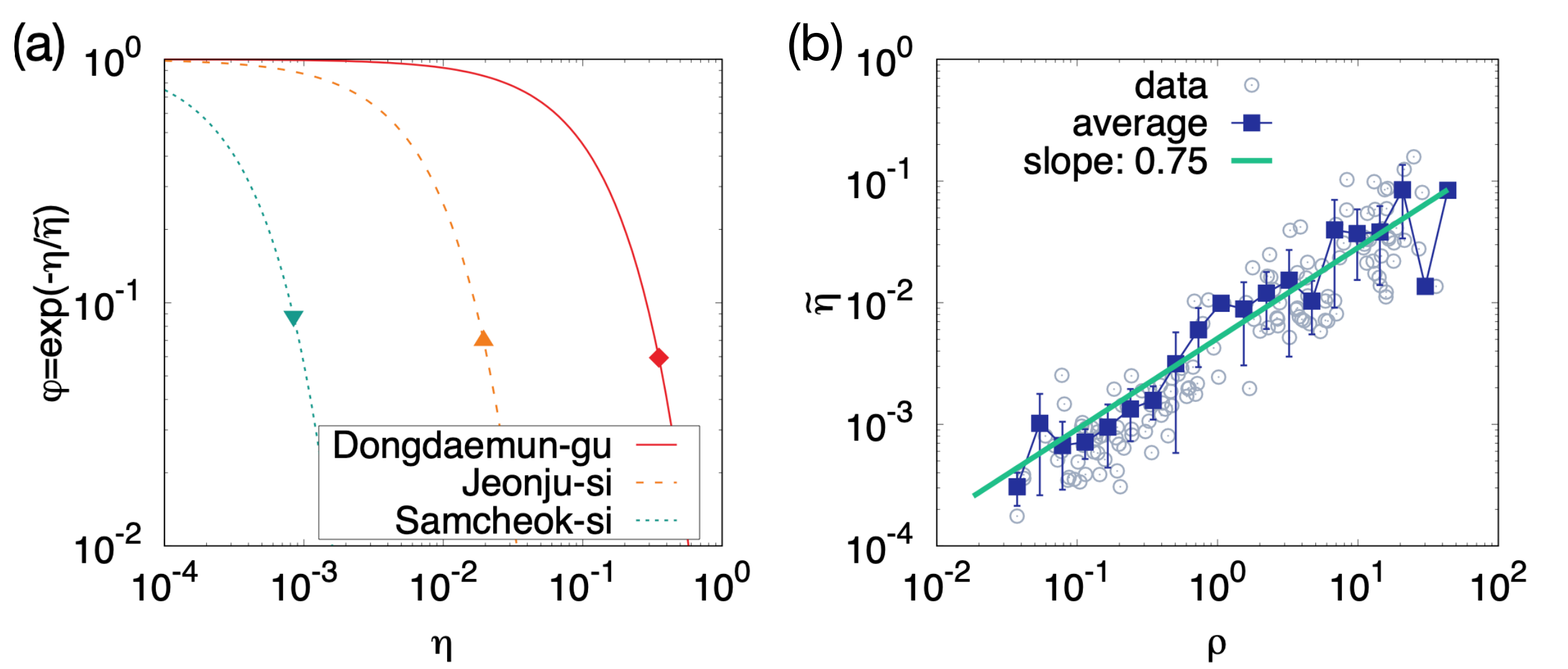}
\centering
\caption{Theoretical prediction for the fatality rate and the estimated characteristic hospital density. 
({\bf a}) The theoretical prediction, Eq.~(\ref{eq:fatality_rate}),  for the fatality rate $\phi$ as a function of the hospital density $\eta$ for selected districts having $\chrh= 1.2\times10^{-1}$, $7.3\times10^{-3}$, and $3.5\times10^{-4}{\rm km}^{-2}$, respectively. Filled points represent the real data for each district. 
({\bf b}) Plot of $\chrh$ versus the patient density $\rho$. Open circles and filled squares indicate the real data and the average, respectively. The solid line fits the average of $\chrh$ as a function of $\rho$ and the slope is $0.75 \pm 0.056$.
}
\label{fig:etatilde}
\end{figure}
%%%%%%%%%%%%%%%%%%%%%%%%%%%%%%%%%%%%%%%%%

Assuming the validity of Eq.~(\ref{eq:fatality_rate}), one can estimate the characteristic hospital density $\chrh_i$ by using the empirical data of the fatality rate $\phi_i$ and the hospital density $\eta_i$  in Eq.~(\ref{eq:fatality_rate}) as 
\begin{equation}
\chrh_i=\frac{\eta_i}{\left| \log \phi_i \right|}.
\label{eq:emp_etatilde}
\end{equation}
The exponential functions $\phi_i(\eta)$'s in Eq.~(\ref{eq:fatality_rate}) with the estimated coefficient $\chrh_i$ for selected districts are shown  in Fig.~\ref{fig:etatilde}(a). $\chrh_i$ is different from district to district, growing with the patient density [Fig.~\ref{fig:etatilde}(b)], which underlies the weaker decay of the fatality rate with the  hospital density [Fig.~\ref{fig:relations}(b)] than would be expected if $\chrh_i$ were identical for all districts. The estimated $\chrh_i$ is the characteristic constant of each district and will be  used throughout the optimization computation. 

%%%%%%%%%%%%% Figure 3: Dimensionless quantities  %%%%
\begin{figure}
\includegraphics[width=1.0\linewidth]{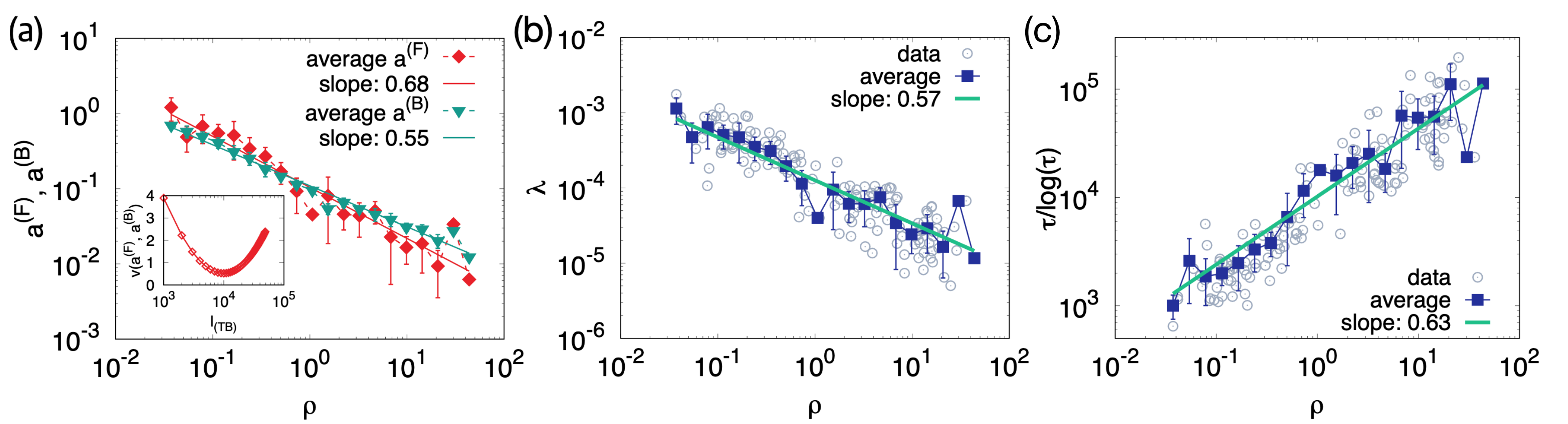}
\centering
\caption{ Lattice constant and dimensionless quantities. 
({\bf a}) Lattice constants $\af = \af(\ltb^*)$ and $a^{(\rm B)}$ as functions of the patient density $\rho$ in logarithmic scales. The errorbars are standard deviations. The fitting lines have slopes $-0.68 \pm 0.044$ and $-0.55 \pm 0.019$, respectively. Inset: The average  logarithmic distance $v$ is minimized at $\ltb^* = 10000$ km with errorbar $1000$ km. 
({\bf b}) Dimensionless hospital density $\lambda=\eta a^2$ [Eq.~(\ref{eq:dimensionless_lambda})] versus patient density. The dashed line with filled squares represents the average values with the errorbars being standard deviations. The solid line fits the average values and has slope $-0.57 \pm 0.050$. 
({\bf c}) Plot of ${\tau \over \log\tau}$ versus patient density.  The dashed line with filled squares and errorbars represent the average values and standard deviations. The slope of the solid line is $0.63 \pm 0.046$.
}
\label{fig:rwmodel}
\end{figure}
%%%%%%%%%%%%%%%%%%%%%%%%%%%%%%%%%%

The lattice constant can be obtained by using the estimated $\chrh_i$ in Eq.~(\ref{eq:etatilde}) and solving for $a_i$.  Let us denote the solution by $\af_i (\ltb)$. To validate it, we compare it with another estimate independent of the empirical values of the fatality rate or the hospital density. We use the data of the  number of business buildings $B_i$ in each district~\cite{kosis}. The business buildings, including hospitals, are the candidates for the notable places included  in the coarse-grained trajectories. The typical distance between adjacent business buildings can be a candidate for the lattice constant, which is given by 
\begin{equation}
a^{(\rm B)}_i = \sqrt{\frac{A_i}{B_i}}
\label{eq:ab}
\end{equation}
under the assumption that the business buildings are uniformly distributed in each district. For the comparison of $a^{\rm (F)}(\ltb)$ and $a_i^{\rm (B)}$, we take the value of $\ltb$ minimizing the average logarithmic distance $v(\af,\ab) =\sum_i ( \log \af_i - \log a^{(\rm B)} )^2/\sum_i 1$, which is $\ltb^* = 10000 \pm 1000$ [Fig.~\ref{fig:rwmodel}(a)]. It corresponds to the annual traveling distance $3300$ km which is reasonably close to the empirical value $8478$ km of Korea~\cite{ktdb}.  In Fig.~\ref{fig:rwmodel}(a), the two lattice constants  $\af = \af(\ltb^*)$ and $a^{(\rm B)}$ show good agreement in their magnitudes, supporting the validity of the assumptions of our model and its formulas,  Eqs.~(\ref{eq:fatality_rate}) and~(\ref{eq:etatilde}). Due to this agreement and Eq.~(\ref{eq:ab}),  we can see that a large or small value of $\af_i$ originates from the sparse or dense business buildings in district $i$. 

With $\af_i$, the dimensionless hospital density $\lambda_i$ and the number of steps $\tau_i$ taken for $\tb$ can be evaluated by Eqs.~(\ref{eq:dimensionless_lambda}) and (\ref{eq:dimensionless_tau}), which are plotted versus the patient population density  in Figs.~\ref{fig:rwmodel}(b) and~\ref{fig:rwmodel}(c), respectively. In contrast to the real hospital density $\eta_i$, $\lambda_i$  is lower in a district with higher patient density [Fig.~\ref{fig:rwmodel}(b)]. It is attributed to the smaller lattice constants in the districts of higher patient densities, arising from the denser buildings. On the other hand, the number of steps taken for $\tb$ increases as the patient density increases, increasing the chance to visit hospitals. To sum up, effectively less hospitals are distributed  but the patients  take more steps for the given period $\tb$ in the higher-populated districts, which explains the slightly lower fatality rates therein than in lower-populated districts as shown in Fig.~\ref{fig:relations} (b). 
%%%%%%%%%%%%%%%%%%%%%%%%%%%%%%%%%%%%%%%%%%%%%%%%%%%%%%%

%%%%%%%%%%%%%%%%%%%%%%%%%%%%%%%%%%%%%%%%%%%%%%%%%%%%%%%
\subsection*{Optimal hospital density}
\label{subsec:optimization}

%%%%%%%%%%%%%% Figure 4 : Optimization %%%%%%%%
\begin{figure}
\includegraphics[width=0.8\linewidth]{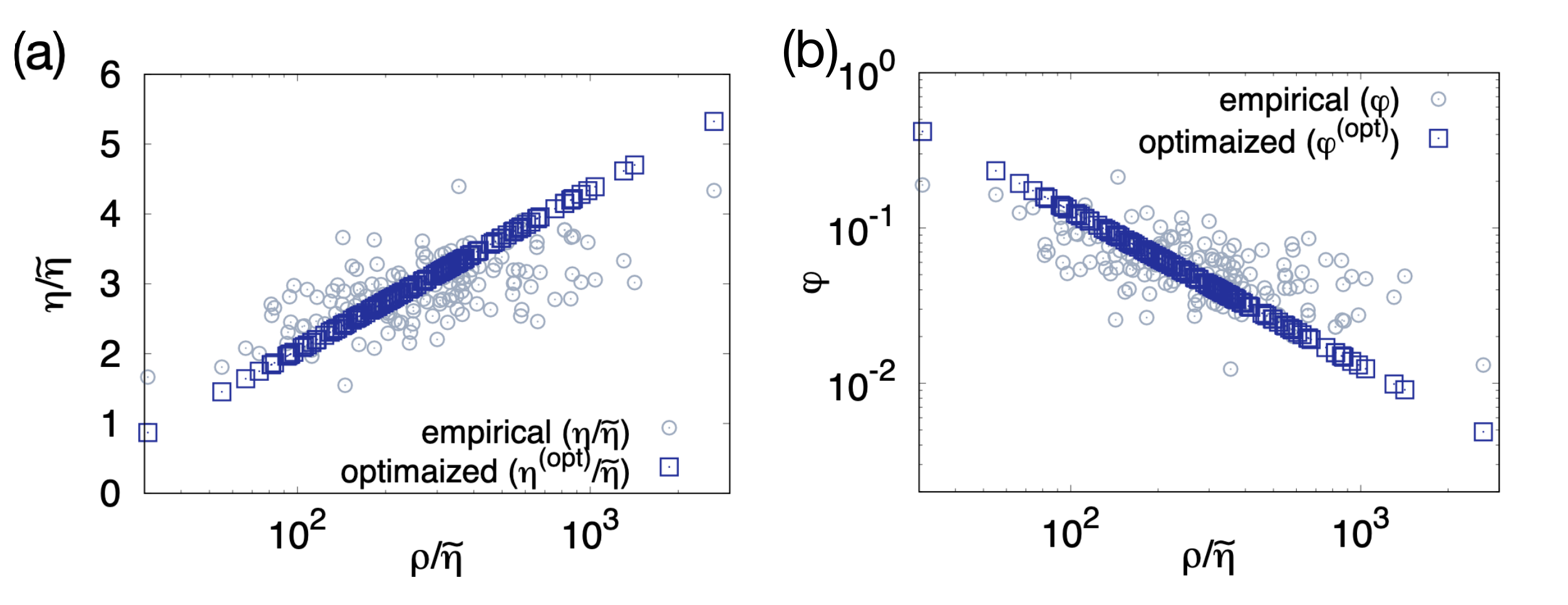}
\centering
\caption{The rescaled hospital density and fatality rate before and after optimization as functions of the rescaled patient density. 
({\bf a}) Plots of the rescaled hospital density, $\eta \over\chrh$ (circle) and $\opth \over \chrh$ (square) versus the rescaled patient density $\rho \over \chrh$ in semilogarithmic scale. $\chrh$ is the characteristic hospital density estimated empirically as Eq.~(\ref{eq:emp_etatilde}). The data points for the optimized hospital density lie on the line corresponding to Eq.~(\ref{eq:opth}) with $z = 12.9$. 
({\bf b}) Plots of the fatality rate $\phi$ (circle) and $\optphi$ (square) versus the rescaled patient density $\rho \over \chrh$. The data points for the optimal fatality rates are on the line corresponding to Eq.~(\ref{eq:optphi}).
}
\label{fig:optimal}
\end{figure}
%%%%%%%%%%%%%%%%%%%%%%%%%%%%%%%%%%%%

The fatality rate formula in Eq.~(\ref{eq:fatality_rate}) is applicable to $I_s$ districts having non-zero $\eta$ and $\phi$ in the empirical data.  Then one can minimize the total fatalities in those $I_s$ districts 
\begin{equation}
E_{\rm fatalities} =\sum_{i=1}^{I_s} N_i \phi_i=\sum_{i} A_i \, \rho_i \, \exp \left(-{ \eta_i \over \chrh_i}\right), 
\label{eq:energy}
\end{equation}
with respect to the hospital density distribution $\{\eta_i\}$ for fixed $N_i$, $\tilde{\eta}_i$, and total number of hospitals $H_{\rm total}$.  In  the data of year 2014, $I_s=143$, $H_{\rm total}=328$, and $E_{\rm fatalities} = 1718$~\cite{kosis}. We allow $H_i$'s to be arbitrary real numbers, and the case of integer $H_i$'s will be discussed later. $E_{\rm fatalities}$ in Eq.~(\ref{eq:energy})  is minimized when $\delta E = \sum_i A_i \delta\eta_i \left(-{\rho_i \over \chrh_i} e^{{-\eta_i \over \chrh_i}} +z\right)=0$ and $\partial^2 E_{\rm fatalities}/\partial \eta_i^2 >0$ with  $z$ being the Lagrange multiplier. Consequently the optimal hospital density is found to be 
\begin{equation}
\opth_i = \chrh_i \log  \left( \frac{\rho_i}{z\chrh_i}\right),
\label{eq:opth}
\end{equation}
and the optimal fatality rate is 
\begin{equation}
\optphi_i = z \frac{\chrh_i}{\rho_i}.
\label{eq:optphi}
\end{equation}
The Lagrange multiplier $z$ is computed by inserting Eq.~(\ref{eq:opth}) into  Eq.~(\ref{eq:Htotal}) as
$z=\exp\left[\frac{\sum_i A_i \chrh_i \log \left( \frac{\rho_i}{\chrh_i} \right) - H^{(\rm total)}}{\sum_i A_i \chrh_i }\right]\simeq 12.9$. 

Equations~(\ref{eq:opth}) and (\ref{eq:optphi}) are the main results of the present study. Remarkably the optimal hospital density and the patient density are rescaled commonly by $\tilde{\eta}_i$  and then related to each other logarithmically. In Fig.~\ref{fig:optimal}(a), the arrangement of the data points of the optimal hospital densities on a straight line is contrasted with the scattered distribution of the current (empirical) hospital densities in the ($\rho/\chrh, \eta/\chrh$) plane  in semi-logarithmic scale. The same phenomenon is observed for the fatality rate; the empirical fatality rates $\phi_i$'s are scattered but the optimized fatality rates  lie on a straight line in the ($\rho/\chrh, \phi$) plane in logarithmic scale as shown in Fig.~\ref{fig:optimal}(b).  The rescaled patient density ranges between 30.80 (Yeonggwang-gun) and 2646 (Songpa-gu), and  is larger than $z\simeq 12.9$ and thus guarantees $\opth_i>0$ for all  $i$ in Eq.~(\ref{eq:opth}).   More plots of the optimized hospital densities and fatality rates are given in Fig.~\ref{fig:scal_opt}.

%%%%%%%%%%% Figure 5 : Simulation %%%%%%%%%%
\begin{figure}
\includegraphics[width=0.8\linewidth]{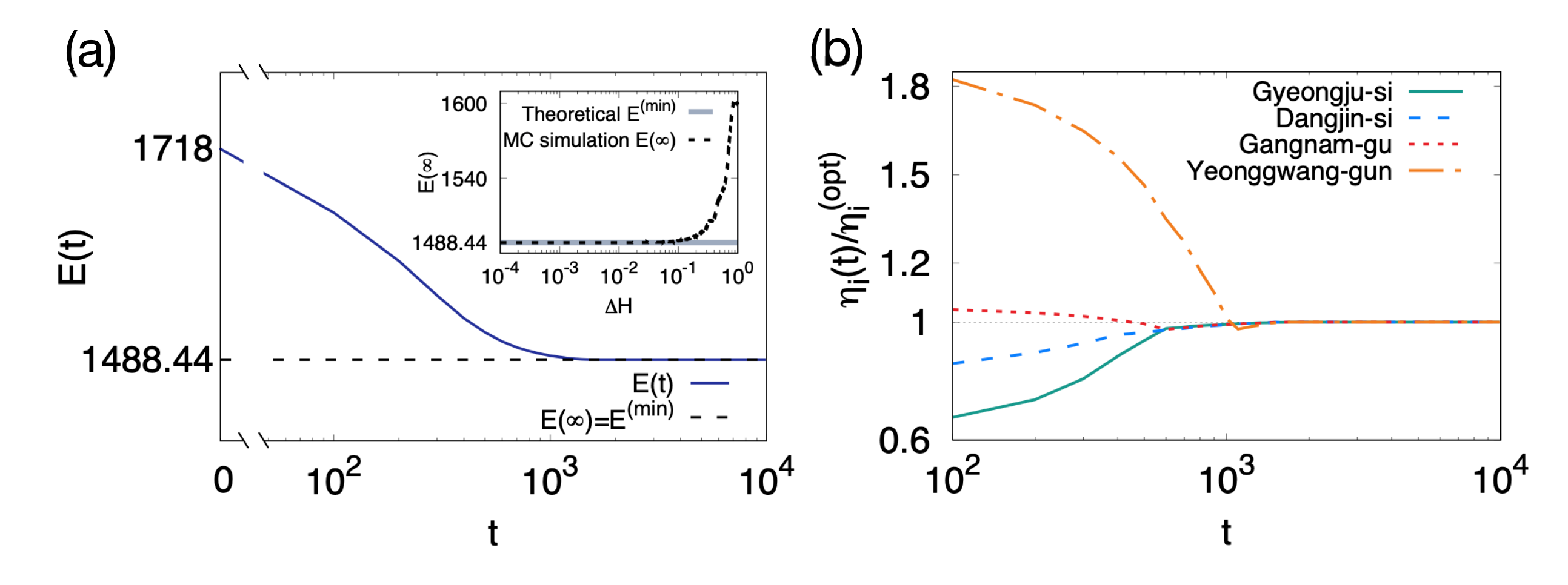}
\centering
\caption{Monte-Carlo (MC) simulation for optimizing the hospital distribution. 
({\bf a}) The total energy $E(t) = E_{\rm fatalities}$ as a function of the MC step $t$ in the MC simulation with $\Delta H=10^{-2}$. It becomes stationary at $E(\infty)= E^{\rm (min)} =1488.44$ for $t\gtrsim 10^3$. Inset: The stationary-state value $E(\infty)$ depends on the increment $\Delta H$. 
({\bf b}) The ratio $\eta_i(t) \over \opth_i$ is plotted as a function of the MC step $t$ for selected districts with $\Delta H=10^{-2}$.  It converges to one for $t\gtrsim 10^3$.
}
\label{fig:simulation}
\end{figure}
%%%%%%%%%%%%%%%%%%%%%%%%%%%%%%%%

The scattered distributions of the empirical data in Fig.~\ref{fig:optimal} show clearly the deviation of the current distribution of hospitals from the optimum minimizing the total fatalities of TB. The minimized total fatalities $E^{(\rm min)}_{\rm fatalities}$ obtained from  the optimal hospital distribution is 
\begin{equation}
E^{(\rm min)}_{\rm fatalities}=\sum_i N_i \frac{z\, \chrh_i}{\rho_i}=z\sum_i A_i \chrh_i \simeq 1488.44,
\label{eq:minE}
\end{equation}
which is smaller than the current value, 1718, by $13$\%. For this optimization,  $\sum_i A_i |\opth_i - \eta_i|/2\simeq 24.4$ hospitals among a total of $H_{\rm total}=328$ are relocated. In the zero-temperature Monte Carlo (MC) simulation in which some small amount $\Delta H$ of hospitals are moved between randomly selected districts only when the attempted relocation reduces $E_{\rm fatalities}$, the theoretical predictions in Eqs.~(\ref{eq:opth}) and (\ref{eq:minE}) are realized in the steady state  as long as  $\Delta H$  is sufficiently small [Fig.~\ref{fig:simulation}(a)]. If $H_i$'s are restricted to be integers ($\Delta H=1$), the total fatalities in the steady state is 1599.45. The stationary-state results remain unchanged in the simulations with different initial configurations or  with gradually cooling down the temperature. For more details of the simulations, see Methods. 

%%%%%%%%%%%%% Figure: Effects of optimization %%%%%%
\begin{figure}
\includegraphics[width=1.0\linewidth]{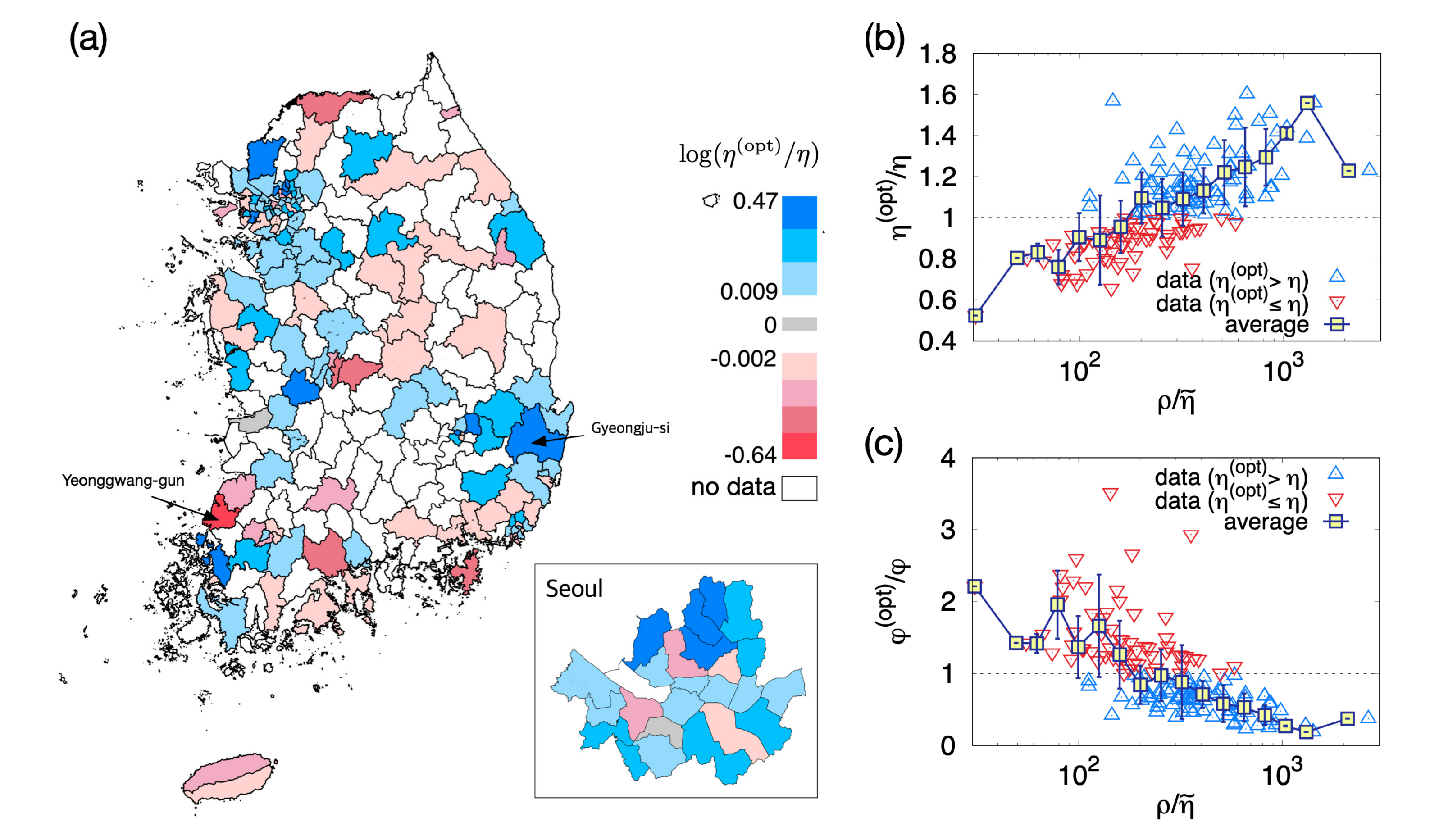}
\centering
\caption{Changes of the hospital density by optimization. 
({\bf a}) The logarithmic ratio of the optimal to current hospital density $\log \left({\opth \over \eta}\right)$ is encoded by color for each district of Korea. 
85 districts are white, as they do not have both $\opth$ and $\eta$ available. 
({\bf b}) Plot of the ratio  $\opth \over \eta$ versus the rescaled patient density $\rho \over \chrh$. Upper and lower triangles are used  for the data points with the optimal hospital density larger and smaller, respectively, than the current one.  The filled square is the average and the errorbar is the standard deviation of the ratio  $\opth \over \eta$. 
({\bf c}) Plot of the ratio of the optimal to current  fatality rate  $\optphi \over \phi$ versus $\rho \over \chrh$. 
}
\label{fig:effect}
\end{figure}
%%%%%%%%%%%%%%%%%%%%%%%%%%%%%%%%%%%%%

To achieve such reduction in the total fatalities, the hospital density should be increased in some districts and decreased in others  [Fig.~\ref{fig:simulation}(b)]. For instance, Gyeongju-si should have its hospital density 1.61  times larger  than the current hospital density but Yeonggwang-gun 0.534 times larger than the current one.  In Fig.~\ref{fig:effect}(a), 143 districts are colored blue (red) if the optimal hospital density is larger (smaller) than the current hospital density. Interestingly, the ratio ${\eta^{\rm (opt)} \over \eta}$ of the optimal to current hospital density turns out to  depend strongly on the rescaled patient density ${\rho \over \tilde{\eta}}$ [Fig.~\ref{fig:effect} (b)]. It implies that if the rescaled patient density is large in a district and small in another district, it is recommended to move some hospitals from the latter to the former district. The high and low values of $\opth\over \eta$ of Gyeongju-si  and Yeonggwang-gun can be understood in this line, as they have quite different values of $\rho\over \tilde{\eta}$,  662 and 30.8, respectively. Such a significant correlation is absent between ${\opth\over\eta}$ and the raw patient density $\rho$ [Fig.~\ref{fig:ratio_rho}]. The change of the fatality rate shows the opposite trend to that of the hospital density; The districts with large (small) rescaled patient density ${\rho\over \tilde{\eta}}$ have their fatality rate decreased (increased) as they gain (lose) hospitals [Fig.~\ref{fig:effect}(c)]. 
%%%%%%%%%%%%%%%%%%%%%%%%%%%%%%%%%%%%%%%%%%%%%%%%%%%%%%%

%%%%%%%%%%%%%%%%%%%%%%%%%%%%%%%%%%%%%%%%%%%%%%%%%%%%%%%
\section*{Summary and Discussion}
\label{sec:summary}

We have here proposed a modeling framework which  predicts the optimal distribution under a general objective function, going beyond the previous descriptive explanations for the facility distribution. In deriving the optimal hospital distribution over districts for minimizing the TB fatalities in the whole country, we have found that the characteristic hospital density of each district plays an important role in the optimization. The random-walk nature of the coarse-grained trajectories of patients has been assumed in establishing a theoretical model,  and the lattice constant of each district has been introduced  to connect the theoretical results and the empirical data. The incorporation of such heterogeneity of districts in the theoretical study of the facility optimization is done only in the present study and can be useful in future studies. 

Examining the assumptions and limitations of the proposed model may help better understand and improve its predictive power. The exponential decay of the fatality rate with  the hospital density given in Eq.~(\ref{eq:fatality_rate}) is valid when the dimensionless hospital density is low, ${\sqrt{\lambda \tau} \over \log \tau}\ll 1$~\cite{rw1}. The empirical data analyzed in the present study stay in this regime; ${\sqrt{\lambda \tau} \over \log \tau}$ ranges between 0.23 and 0.37. If we were to extend to the case of high hospital density or the hospital locations being no more independent of one another, the fatality rate might behave differently from Eq.~(\ref{eq:fatality_rate}). For $\phi_i = f(\eta_i/\tilde{\eta}_i)$ with $f(x)$ a decreasing convex function such as exponential, stretched exponential, or power law,  the optimal hospital density will be given by ${\eta_i \over \chrh_i} = (-f')^{-1} \left(z { \chrh_i \over \rho_i}\right)$ with $(f')^{-1}$ being the inverse of the derivative of $f(x)$.  The relation between the two rescaled variables ${\eta_i \over \chrh_i}$ and ${\rho_i \over \chrh_i}$ depends on the specific form of $f(x)$ and reduces to Eq.~(\ref{eq:opth}) in case of $f(x)=e^{-x}$. The case of $f(x)$ being a power law is presented in the Supplementary Information (SI). Regarding the robustness of the functional form of the fatality rate,  it will be of interest to investigate which model of walk with traps exhibits such a power-law survival probability.

We have counted only the private general hospitals, but  there are mostly found one or two public health centers in a district, which can also provide the medical treatment to TB patients although its portion may not be large~\cite{lee2015overview}. One can optimize the private hospital distribution considering the contribution of the public health centers to the TB treatment, which is presented in the SI and Fig.~\ref{fig:all}. The results remain unchanged qualitatively. Extending the trajectories of patients to the nearby districts can be one way of making the model more realistic, which will address how the similarity or dissimilarity of adjacent districts may affect the fatality rate and the total fatalities. When a given number of hospitals can be opened or should be shut down for financial or other reasons, our results will be helpful for the investigation of the optimal locations. 
%%%%%%%%%%%%%%%%%%%%%%%%%%%%%%%%%%%%%%%%%%%%%%%%%%%%%%%

%%%%%%%%%%%%%%%%%%%%%%%%%%%%%%%%%%%%%%%%%%%%%%%%%%%%%%%
\section*{Methods}
\subsection*{Data-sets}
Under the control of the Ministry of Health and Welfare, several organizations such as the Korean National Tuberculosis Association and Korea Centers for Disease Control and Prevention cooperate to prevent and eradicate tuberculosis in Korea~\cite{mohw}. The related information has been well recorded, which is accessible through Statistics Korea~\cite{kosis}. The data-sets used in the present study have been collected district by district. As a result, 37347 new TB patients, 330 hospitals, and 2127 dead patients for 228 districts in year 2014 have been considered in the present study. The hospitals considered in our study are the private ones classified as general or superior general hospitals in the Korean Medical Service Act.  The theoretical modeling for the fatality rate applies for  $I_s=143$ districts which have at least one hospital and at least one dead patient. The total number of new and dead patients, and hospitals in those 143 districts are 32322, 1718, and 328.

\subsection*{Monte Carlo simulation}
To illustrate the hospital relocation process, we perform the zero-temperature Monte Carlo (MC) simulation in which hospitals are relocated over  $I_s$ districts towards decreasing the energy, equal to the total fatalities given in Eq.~(\ref{eq:minE}), as follows:
\begin{enumerate}[label=(\roman*)]
\item Initially the number of hospitals in each district is set equal to the empirical data. 
\item For two randomly selected districts $i$ and $j$, consider moving $\Delta H$ hospitals from $i$ to $j$ as long as $H_{i}-\Delta H>0$.  
\item Accept this relocation if  the energy change  $\Delta E = N_i (e^{-{H_i - \Delta H \over A_i \chrh_i }}  -e^{-{H_i  \over A_i \chrh_i }}  ) + N_j (e^{-{H_j + \Delta H \over A_j \chrh_j }}  -e^{-{H_j  \over A_j \chrh_j}} )$ is zero or negative. Reject it otherwise.  
\item Repeat steps (ii) and (iii) $I_s$ times to increase the MC step $t$ by one.  
\end{enumerate}
We find that the energy becomes stationary around $t=10^3$ MC steps and thus we run the simulations just up to $10^4$ MC steps [Fig.~\ref{fig:simulation}(b)]. $\Delta H$ represents the amount of hospitals moved by one relocation.  For $\Delta H\lesssim 0.05$,  the hospital configuration and the energy in the stationary state coincide with the theoretical predictions in Eqs.~(\ref{eq:opth}) and~(\ref{eq:minE}), respectively.  Replacing the initial hospital configuration by a random one, the energy and the hospital configuration in the stationary-state are not changed but remain the same as the theoretical prediction. Since a hospital relocation is accepted only when the corresponding energy change is not positive, this simulation corresponds to zero temperature $T=0$. We have also run the MC simulation with lowering temperature gradually from $T=20$ to $T=3\times 10^{-9}$ but the hospital configuration and the energy in the stationary state are found to be the same as those of the zero-temperature MC simulation. 
%%%%%%%%%%%%%%%%%%%%%%%%%%%%%%%%%%%%%%%%%%%%%%%%%%%%%%%

\section*{Data availability}
The datasets generated during  the current study are available from the corresponding author on reasonable request.

%%%%%%%%%%%%%%%%%%%%%%%%%%%%%%%%%%%%%%%%%%%%%%%%%%%%%%%
%%% References
%%%%%%%%%%%%%%%%%%%%%%%%%%%%%%%%%%%%%%%%%%%%%%%%%%%%%%%
%\bibliography{Hospital}

%%%%%%%%%%%%%%%%%%%%%%%%%%%%%%%%%%%%%%%%%%%%%%%%%%%%%%%
%%% Acknowledgements
%%%%%%%%%%%%%%%%%%%%%%%%%%%%%%%%%%%%%%%%%%%%%%%%%%%%%%%

\section*{Acknowledgements}
This work was supported by the National Research Foundation of Korea (NRF) grants funded by the Korean Government (No. 2019R1A2C1003486).

%%%%%%%%%%%%%%%%%%%%%%%%%%%%%%%%%%%%%%%%%%%%%%%%%%%%%%%
%%% Author contribution
%%%%%%%%%%%%%%%%%%%%%%%%%%%%%%%%%%%%%%%%%%%%%%%%%%%%%%%
\section*{Author Contributions}
M.J.L. analyzed the data-sets, developed the theory, performed the simulations, and wrote the manuscript. K. K. and J. S analyzed the data-sets. D.-S.L. designed and supervised the research, and wrote the  manuscript.

\section*{Competing Interests}
The authors declare no competing interests.

%%%%%%%%%%%%%%%%%%%%%%%%%%%%%%%%%%%%%%%%%%%%%%%%%%%%%%%
%%% Supplementary Information
%%%%%%%%%%%%%%%%%%%%%%%%%%%%%%%%%%%%%%%%%%%%%%%%%%%%%%%

\appendix
\section*{\center \LARGE Supplementary Information}
\setcounter{figure}{0}    
\setcounter{equation}{0}
\renewcommand{\thesection}{S\arabic{section}}
\renewcommand{\theequation}{S\arabic{equation}}
\renewcommand{\thefigure}{S\arabic{figure}}
\renewcommand{\thetable}{S\arabic{table}}
\newcommand{\prvh}{\eta^{({\rm private})}}
\newcommand{\pubh}{\eta^{({\rm public})}}

%%%%%%%%% Figure S1 : Scaling  wrt patient density
\begin{figure}
\includegraphics[width=\linewidth]{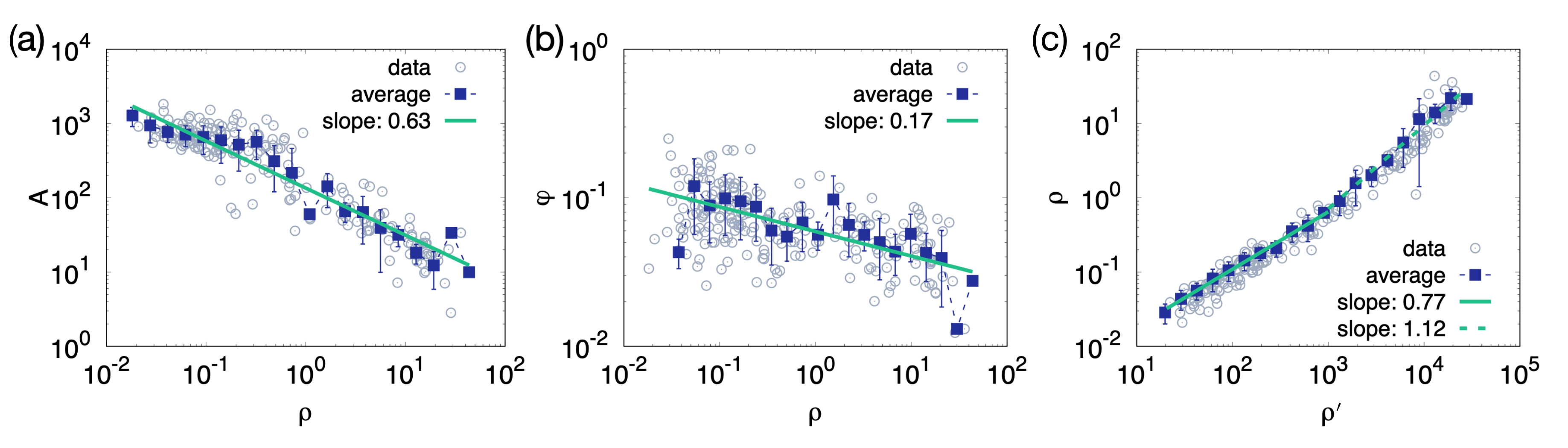}
\centering
\caption{ 
Scaling behaviors of the properties of districts. 
({\bf a}) Plot of the area $A$ versus the patient density $\rho $. 
({\bf b}) Plot of the fatality rate $\phi$ versus the patient density $\rho$.
({\bf c}) Plot of the patient density $\rho$ and the whole population density $\rho'$.
In all the panels,  the empirical data points (open circles) and the average values (filled squares) as functions of $\rho$ or $\rho'$ are presented. The errorbar is the standard deviation. In the panel ({\bf c}), the two solid lines having slopes 0.77 and 1.12 fit the average values in the range $\rho'\leq 10^3$ and $\rho'>10^3$, respectively.
}
\label{fig:scal_other}
\end{figure}
%%%%%%%%%%%%%%%%%%%%%%%%%%%%%%%%%%%%%%%%

%%%%%%%%%%% Figure S2: optimal hospital density %%%%%%%%
\begin{figure}
\includegraphics[width=0.8\linewidth]{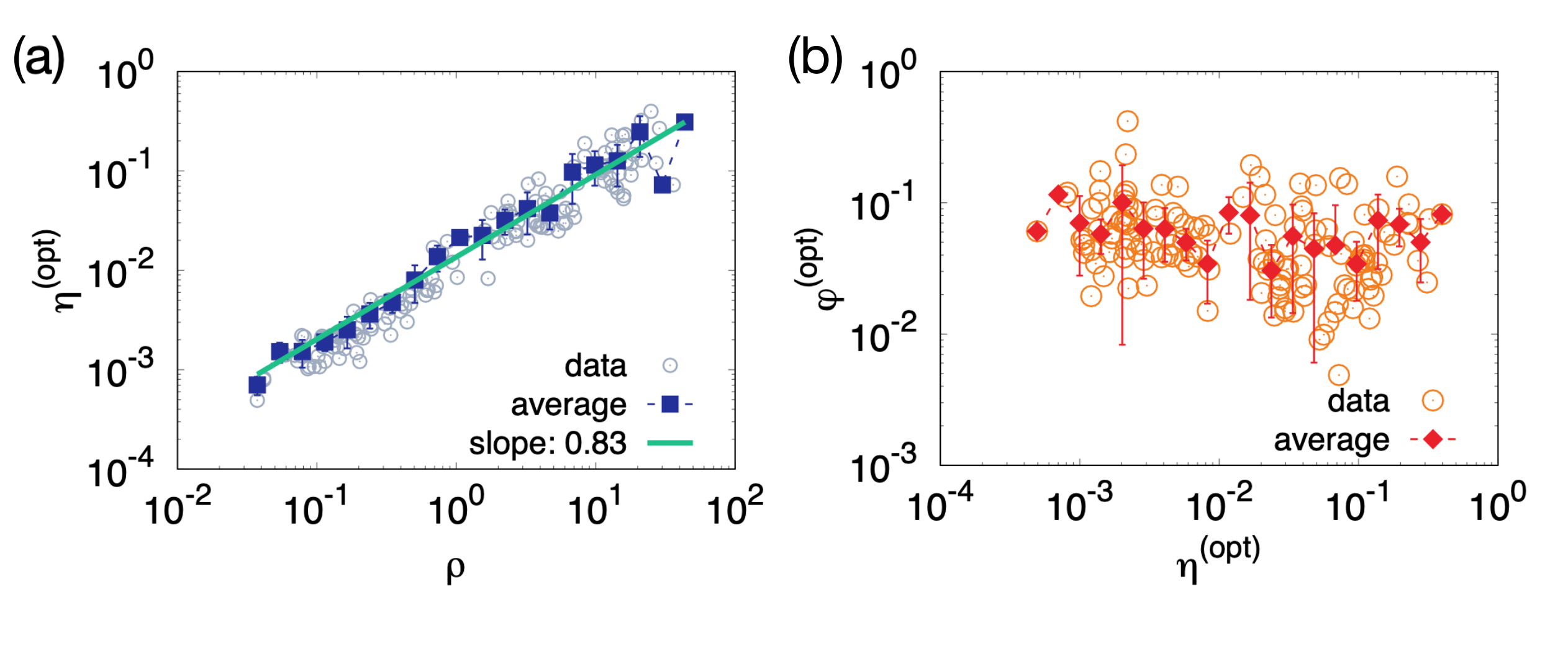}
\centering
\caption{Behaviors of the optimized hospital density and fatality rates. 
({\bf a}) Plot of the optimal hospital density $\opth$ versus the patient density $\rho$. The filled square is the average and the errorbar is the standard deviation. The solid line fits the average of $\opth$ as a function of $\rho$. 
({\bf b}) Plot of the optimal fatality rate $\optphi$ versus the optimal hospital density $\opth$. The filled diamond is the average and the errorbar is the standard deviation.
}
\label{fig:scal_opt}
\end{figure}
%%%%%%%%%%%%%%%%%%%%%%%%%%%%%%%%%%%%

%%%%%%%%%%% Figure S3: eta ratio versus raw rho %%%%%%%%%%
\begin{figure}
\includegraphics[width=0.8\linewidth]{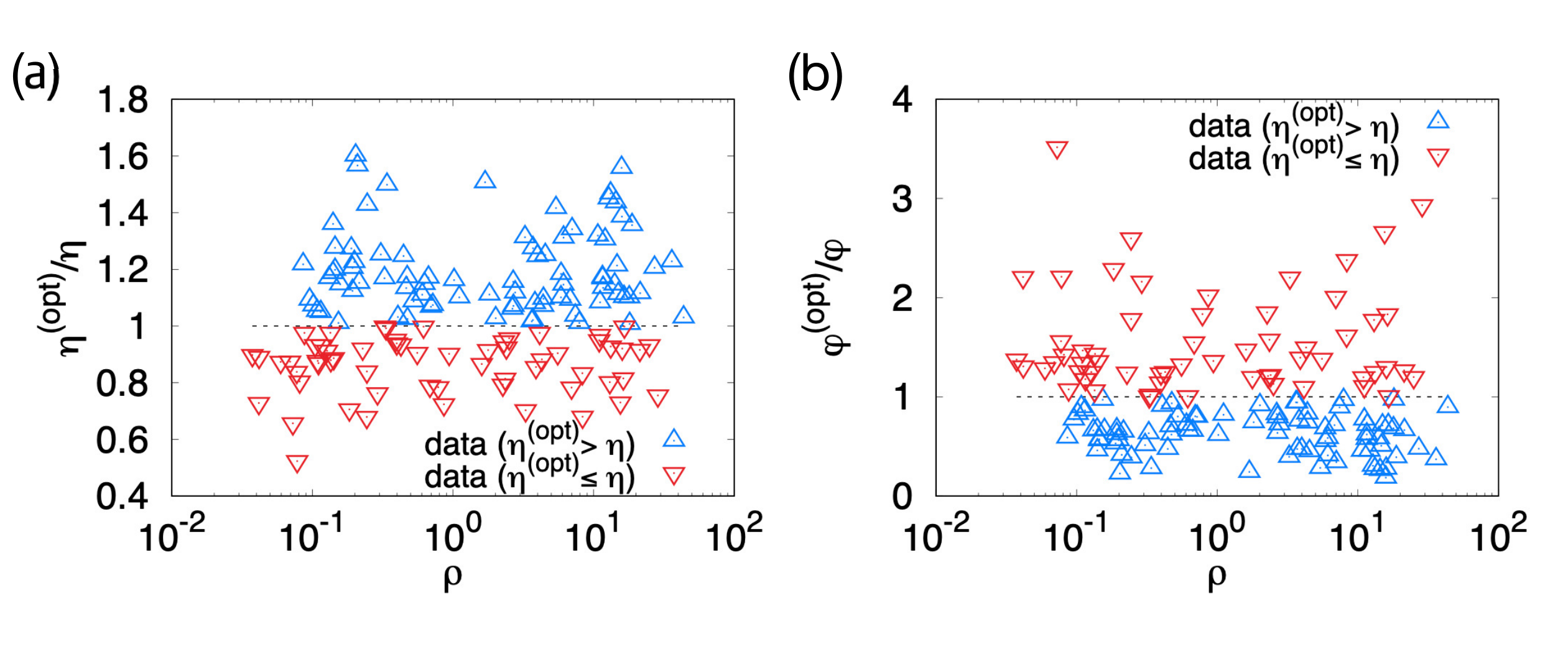}
\centering
\caption{The relation between the changes made by optimization and the raw patient density. 
({\bf a}) Plot of the ratio of the optimal to current hospital density $\opth \over \eta$ versus the raw patient density $\rho$. The data points  are blue (red) for $\opth> \eta$ (for $\opth<\eta$). The data points are scattered, showing no significant correlation. 
({\bf b}) Plot of the ratio of the optimal to current fatality rate $\optphi \over \phi$ versus the raw patient density $\rho$. No correlation is seen.
}
\label{fig:ratio_rho}
\end{figure}
%%%%%%%%%%%%%%%%%%%%%%%%%%%%%%%%

%%%%%%%%%%%%%%%%%%%%%%%%%%%%%%%%%%%%%%%%%%%%%%%%%%%%%%%

\subsection*{Optimizing hospital density under the fatality rate in a power-law form}
\label{subsec:si_plw}

Here we investigate the optimal hospital density in case of $\phi(\eta)$ given as a power law. Suppose that the fatality rate $\phi_i(\eta_i)$ takes the form 
\begin{equation}
\phi_i (\eta_i ) = f\left({\eta_i \over \tilde{\eta}_i}\right),
\label{eq:phi_general}
\end{equation}
with $\chrh_i$ the characteristics hospital density of district $i$ and $f(0)=1$.  Then the total fatalities in Eq.~(\ref{eq:Efatalities}) with $\phi_i$ in Eq.~(\ref{eq:phi_general}) is minimized when $\delta E = \sum_i A_i \delta \eta_i \left( \rho_i {\partial\phi_i \over {\partial \eta_i}}+z\right)=0$ is satisfied or
\begin{equation}
-f'\left({\eta_i \over \chrh_i} \right) = z {\chrh_i \over \rho_i}.
\label{eq:si_opth}
\end{equation}

To be specific, let us consider the fatality rate given in the following power-law form:
\begin{equation}
\phi_i = \left( 1+\frac{\eta_i}{\chrh_i}\right)^{-\gamma}
\label{eq:si_phi}
\end{equation}
with $\gamma$ a constant. By Eq.~(\ref{eq:si_opth}),  the optimal hospital density is determined as $\gamma \left( 1+\frac{\opth_i}{\chrh_i} \right)^{-\gamma-1}=z {\chrh_i \over \rho_i}$, leading to 
\begin{equation}
{\opth_i \over \chrh_i} = \left({\gamma \rho_i \over z \chrh_i }\right)^{1\over \gamma+1} - 1.
\label{eq:si_finalopth}
\end{equation}
The Lagrange multiplier $z$ is determined by the constraint in Eq.~(\ref{eq:Htotal}) and evaluated in this case as 
\begin{equation}
z^{1 \over {\gamma+1}}=\frac{\sum_i A_i\chrh_i \left( {\gamma\rho_i \over \chrh_i}\right)^{1 \over {\gamma+1}} }{H^{\rm (total)}+\sum_i A_i\chrh_i },
\label{eq:si_z}
\end{equation}
and the optimal fatality rate is
\begin{equation}
\optphi_i = \left( \frac{z\, \chrh_i}{\gamma\rho_i}\right)^{\gamma \over {\gamma+1}}.
\label{eq:si_optphi}
\end{equation}
%%%%%%%%%%%%%%%%%%%%%%%%%%%%%%%%%%%%%%%%%%%%%%%%%%%%%%%

%%%%%%%%%%%%%%%%%%%%%%%%%%%%%%%%%%%%%%%%%%%%%%%%%%%%%%%
\subsection*{Analysis including the public health centers}
\label{subsec:si_all}

%%%%%%%%% Figure S4 : Including the public health centers 
\begin{figure}
\includegraphics[width=1\linewidth]{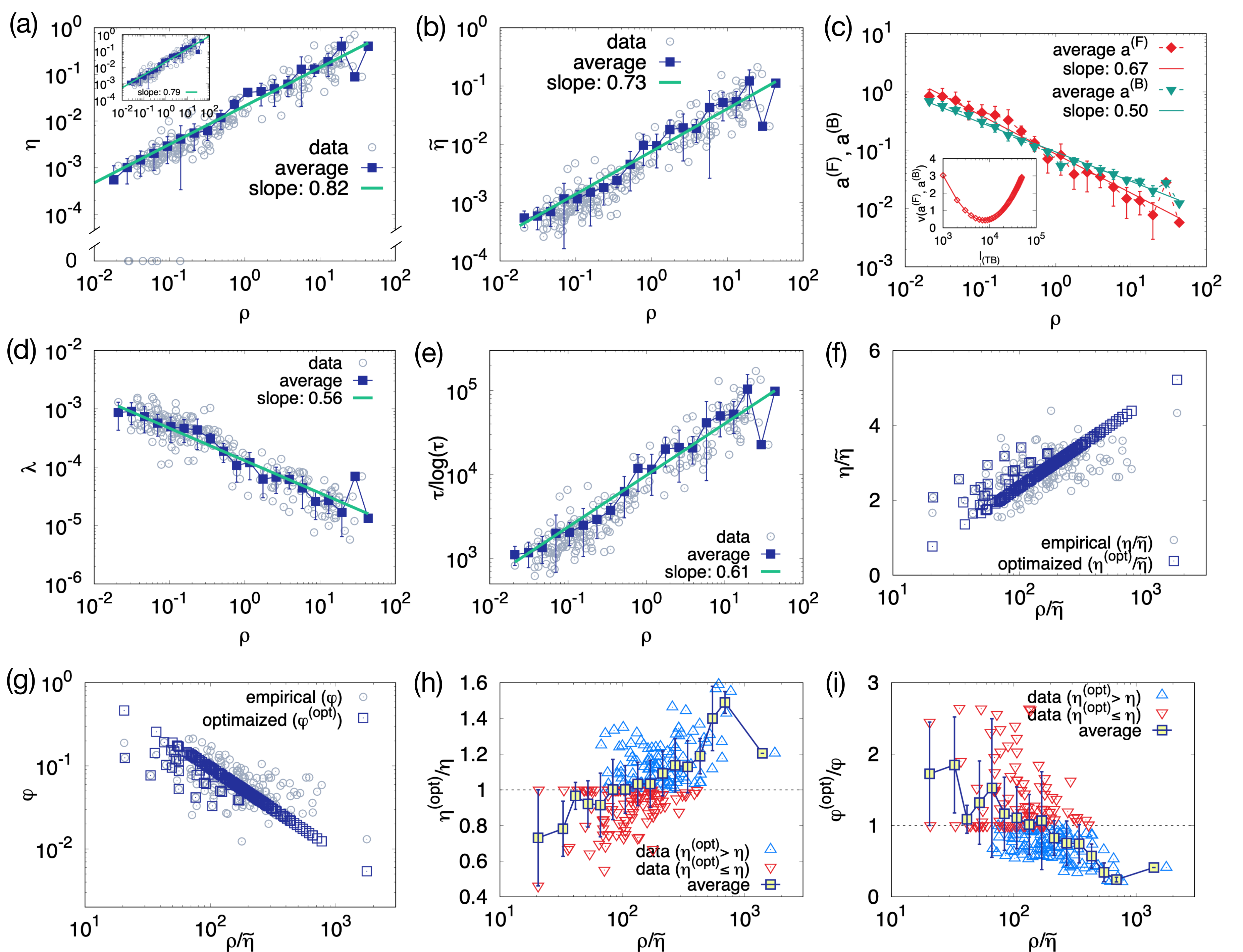}
\centering
\caption{Results of the optimization including the public health centers. 
({\bf a}) Plot of the integrated hospital density $\eta=\prvh+\pubh$  versus the patient density $\rho$. Inset: the same plot for the districts having nonzero $\eta$ and nonzero $\rho$.
({\bf b}) Plot of the integrated characteristic hospital density $\chrh$ versus the patient density $\rho$.
({\bf c}) Plot of the lattice constant $\af$, obtained by using the integrated characteristic density in Eq.~(\ref{eq:etatilde}) with $\ltb^*=8000$ km,  versus the patient density $\rho$, compared with  $a^{\rm (B)}$. Inset: The logarithmic distance between $\af(\ltb)$ and $a^{\rm (B)}$ as a function of $\ltb$. It is the minimum at $\ltb = 8000$ km. 
({\bf d})  Plots of the dimensionless hospital density $\lambda$  versus the patient density $\rho$.
({\bf e}) Plots of $\tau/\log \tau$ versus the patient density $\rho$.
({\bf f}) Plots of the rescaled hospital density $\eta \over \chrh$  before and after optimization versus the rescaled patient density $\rho \over \chrh$. 33 districts have $\opth_i=\pubh_i$, so their optimized data points deviate from  the straight line.
({\bf g}) Plots of the fatality rate $\phi$ before and after optimization versus the rescaled patient density $\rho \over \chrh$. 
({\bf h}) Plots of the ratio of the hospital density before and after optimization $\opth \over \eta$ versus the rescaled patient density $\rho \over \chrh$.
({\bf i}) Plots of the ratio of the fatality rate before and after optimization $\optphi \over \phi$ versus the rescaled patient density $\rho \over \chrh$.
}
\label{fig:all}
\end{figure}
%%%%%%%%%%%%%%%%%%%%%%%%%%%%%%%%

Most districts have one or two public health centers each. They  are also in charge of providing the medical treatment for TB patients. Therefore it may be interesting to incorporate the contribution of public health centers to the fatality rate in the investigation of the optimal distribution of private hospitals.

Let the hospital density of a district be the sum of public and private ones as
\begin{equation}
\eta_i=\prvh_i+\pubh_i\geq \pubh_i,
\label{eq:si_conditions}
\end{equation}
where $\prvh_i$ is equal to the hospital density considered in the main text, defined as the ratio of the number of private hospitals to the area of district $i$, and $\pubh_i$ is the ratio of the number of public health centers to the district's area. The characteristic hospital density $\chrh_i$ is computed by using Eq.~(\ref{eq:si_conditions})  in Eq.~(\ref{eq:emp_etatilde}). These integrated hospital density and characteristic density are plotted as functions of the patient density in Figs.~\ref{fig:all}(a) and~\ref{fig:all}(b), respectively. The scaling exponents are similar to those obtained when only the private hospitals are considered.

The lattice constant $\af$ obtained by using the integrated characteristic density in Eq.~(\ref{eq:etatilde}) with $\ltb^*=8000$ km and $\ab$ are compared as functions of the patient density in Fig.~\ref{fig:all}(c), and the dimensionless hospital density $\lambda$ and the number of steps $\tau$ of each district are given in Fig.~\ref{fig:all}(d) and~\ref{fig:all}(e), respectively. The reasonable agreement of $\af$ and $\ab$, the decrease of $\lambda$ and the increase of $\tau/\log \tau$ with increasing the patient density $\rho$ are observed as in the case of considering the private hospitals only.

Let us consider the relocation of private hospitals, fixing public health centers, across districts.  The hospital density should be equal to or larger than the fixed public health center density, i.e., $\eta_i \geq \pubh_i$. Note that the constraint $\eta_i>0$ is used in the main text where only the private hospitals are considered. For the relocation of private hospitals, we consider 217 districts which have at least one private or public hospital and non-zero fatality rate.  The total number of hospitals in those districts is 568, and the total fatalities is 2108 in the empirical data.

Incorporating the inequality constraint of Eq.~(\ref{eq:si_conditions}) as well as the equality of Eq.~(\ref{eq:Htotal}) in the optimization, we find the Karush-Kuhn-Tucker (KKT) conditions~\cite{kkt1, kkt2, kkt3} in minimizing the total fatalities as 
\begin{equation}
\frac{\delta}{\delta \eta_i} \sum_i \left[ N_i\,\exp\left(-\frac{\eta_i}{\chrh_i}\right) -z\left( H^{(\rm total)}-\sum_i \eta_i \, A_i \right) - w_i\, \left(\eta_i-\pubh_i\right)\right]=0,
\label{eq:si_minimize}
\end{equation}
with $z$ and $w_i$'s called the Lagrange multiplier and the KTT multipliers respectively. 
Then the optimal hospital density, including both public and private, is given by 
\begin{equation}
\opth_i(z,w_i) = \chrh_i \log \left( \frac{1}{z+{w_i\over A_i}}\frac{\rho_i}{\chrh_i}\right), 
\label{eq:si_allopth}
\end{equation}
and the optimal fatality rate is 
\begin{equation}
\phi^{\rm (opt)}_i(z,w_i) = \left(z + {w_i \over A_i}\right) {\chrh_i \over \rho_i}.
\label{eq:si_alloptphi}
\end{equation}
These are reduced to Eqs.~(\ref{eq:opth}) and (\ref{eq:optphi}), respectively, if $w_i=0$.  To meet the inequality and equality conditions, it is known~\cite{kkt1, kkt2, kkt3}  that  the optimal hospital density either satisfies $w_i=0$ or $\eta_i = \pubh_i$ for every $i$. Therefore the true optimal solution $\optphi(z^{\rm (opt)}, w_i^{\rm (opt)})$ can be found practically by finding $z^{\rm (opt)}$ with which  i) the optimal hospital density of every district is given either by Eq.~(\ref{eq:si_allopth}) with $w_i=0$ or by $\opth_i = \pubh_i$, and ii) the total number of hospitals is equal to the empirical value as in Eq.~(\ref{eq:Htotal}). To determine  $z^{\rm (opt)}$ and $\{w_i^{\rm (opt)}\}$, we use the following algorithm: 
\begin{enumerate}[label=(\roman*)]
\item For given $z$, the optimal hospital density $\opth_i(z)$ is determined as follows. First use $w_i=0$ in Eq.~(\ref{eq:si_allopth}) to obtain  $\opth_i(w_i=0,z)$ for every district $i$. If it is equal to or larger  than the public center density $\pubh_i$, then accept it as $\opth_i(z)$. Otherwise, $\opth_i(z)$ is set equal to $\pubh_i$, leaving a negative value of $w_i^{\rm (opt)}$ by Eq.~(\ref{eq:si_allopth}). In summary, $\opth_i(z) =\max\{\opth_i(w_i=0,z), \pubh_i\}$. $w_i^{\rm (opt)}=0$ if the former is chosen and $w_i^{\rm (opt)} = A_i [e^{-{\eta_i^{\rm (public)} \over \chrh_i}} {\rho_i \over \chrh_i} -z] $ otherwise.
\item After running step (i) for all  districts, compute the predicted total number of  hospitals $H_{\rm total}^{\rm (opt)}(z) = \sum_i A_i \opth_i(z)$
\item Repeat steps (i) and (ii) for $z$ between 0 and 30 with increment $0.0001$.
\item Determine $z^{\rm (opt)}$ with which the predicted total number of hospitals is the closest to the empirical value, i.e., $|H_{\rm total}^{\rm (opt)}(z) - H_{\rm total}|$ is minimized. 
\end{enumerate}
We find that the predicted total number of hospitals is closest to the empirical value 568 at $z^{\rm (opt)} = 9.5175$, with which $H_{\rm total}^{\rm (opt)}(z^{\rm (opt)}) =567.9996$, and 33 districts have only public health centers with no private hospital, i.e., $\opth_i=\pubh_i$. The total fatalities is reduced to 1878.48, smaller than the current value 2108 by 11\%. 

The scattered distribution of the empirical data and the line alignment of the optimized hospital density in the $(\rho/\chrh, \eta/\chrh)$ plane are also shown in Fig.~\ref{fig:all}(f). Some points for the optimal hospital density deviate from the aligned line,  which are from the 33 districts having $\opth_i = \pubh_i$ in the optimized state. The scaling relation between the optimized fatality rate and the rescaled patient density predicted by Eq.~(\ref{eq:si_alloptphi}) is also shown in Fig.~\ref{fig:all}(g) with the same kind of deviations as in Fig.~\ref{fig:all}(f). The changes of the hospital density and of the fatality rate by the optimization are correlated  with the rescaled patient density positively and negatively as shown in Figs.~\ref{fig:all}(h) and~\ref{fig:all}(i), respectively. These correlations are identical to those in the case of considering the private hospitals only. All these results suggest that even when including the public health centers, all the analysis results remain the same qualitatively. 
%%%%%%%%%%%%%%%%%%%%%%%%%%%%%%%%%%%%%%%%%%%%%%%%%%%%%%%

\end{document}